# POLARIZATION VERSUS AGGLOMERATION

**Vítor João Pereira Domingues Martinho**

Unidade de I&D do Instituto Politécnico de Viseu
Av. Cor. José Maria Vale de Andrade
Campus Politécnico
3504 - 510 Viseu
**(PORTUGAL)**
e-mail: vdmartinho@esav.ipv.pt

**ABSTRACT**

The aim of this paper is to analyze the processes of polarization and agglomeration, to explain the mechanisms and causes of these phenomena in order to identify similarities and differences. As the main implication of this study should be noted that both process pretend to explain the concentration of economic activity and population in certain places, through cumulative phenomena, but with different perspectives, in other words, the polarization with a view of economic development and agglomeration with a perspective of space.

## 1. INTRODUCTION

Over the course of economic development has seen something that can be designated as the phenomenon of "dualism" and there is, therefore, a clear tendency for social, economic and geographical division between center and periphery, urban and rural and town and countryside, a fact that has heightened and strengthened, especially from World War II.

This process is not primarily the result of inherent differences in the exogenous resources at different locations, as advocated by the neoclassical theorists, driven primarily by the forces of supply of inputs. But above all, the existence of increasing returns and endogenous factors, which underlie the creation of a process of circular and cumulative growth, as emphasized by (1)Myrdal (1957) and (2-4)Kaldor (1966, 70 and 81). However, the ideas of Myrdal were not new, since authors such as (5)Young (1928), Adam Smith and (6)Marshall (1920), among others, had already alerted for the issues of increasing returns and cumulative processes, although with other perspectives.

It was based, then in the idea of increasing returns and circular and cumulative processes that appeared after Myrdal two types of processes, agglomeration and polarization, to try to explain the "dualism", associated with different theoretical approaches.

The agglomeration with concerns of location, has tried to explain, especially where economic activity is located and why, trying thus to explain the geographical dualism. The polarization, in turn, approaches related to economic development, seeks to explain why certain areas are more developed than others, trying to explain the dualism and the forces promoting economic growth and development.

## 2. POLARIZATION VERSUS AGGLOMERATION, SIMILARITIES AND DIFFERENCES

The polarization as a result of cumulative growth, walked always associated with the Keynesians, where the variables associated with the forces of demand have particular importance. The explanation of the process of agglomeration with spatial objectives and concerns, already referred, walked more associated with economic geography, where the variable space assumes especial importance.

The polarization process is widely associated with authors like Myrdal, Hirschman and Kaldor, is inspired, as noted, by the "High Development Theory" and is based on the existence of increasing returns to scale in the industry (almost only in this sector), advantages comparative, endogenous and competitiveness of economies. The relationship of increasing returns with the circular and cumulative process develops, in opposition with the process of agglomeration, through the Verdoorn's law, where the productivity is endogenous and depends on the growth of output. In this way, the economic growth is, for the Keynesian, driven by the exogenous demand, the growth of output depends on the strength of demand, especially demand autonomous (as Kaldor, influenced by Harrod, calls), where the main component is the exports. On the other hand, exports are an endogenous part of the demand and depends on efficiency wages (between wages and productivity) that are lower where productivity is higher and this is where the greatest growth of output, too, is larger and so on. It is then in this way that develop the polarization process, an explanation that was due, above all, of the work of Kaldor. Thirlwall, in turn, added to these explanations the importance of the balance of payments has on economic growth, where a balance of payments in deficit may limit demand and therefore growth. The figure below shows a summary of this process.

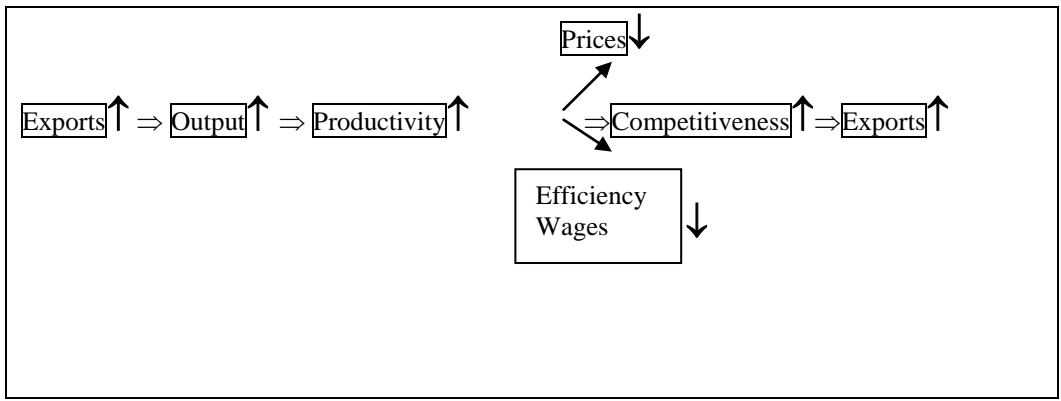

***Figure 1:*** *A mechanism that describes the process of polarization*

The process of agglomeration, in turn, is currently associated with the New Economic Geography, and authors such as Krugman, Venables and Fujita. The main difference this new approach to spatial issues in relation to Traditional Economic Geography, resides primarily in an attempt to reconcile and to model the existence of increasing returns to scale with the market structure of imperfect competition. However, models with the introduction of the variable space, is not exactly an easy task, these authors manage to develop tractable models, have adopted some "tricks" (name them), such as modeling monopolistic competition of the (7)Dixit-Stiglitz (1977 ), the "ad hoc dynamics" and transportation costs, "iceberg". Therefore, the agglomeration process is explained, based also in the inspiration of the "High Development Theory" and taking into account the work of traditional economic geography, especially those associated with known geometry German Weber (1909), (8)Christaller ( 1933) and (9)Losch (1940), the cumulative processes of (10)Pred (1966), the market potential of (11)Harris (1954) and (12)Lowry (1964), the local external economies of Marshall (1920) and (13)Henderson (1974) and (14)von Th$ü$nen model (1826), modeled in conciliation with the market structure, through the "tricks" mentioned above. Considering the above, the process is described through "backward and forward" linkages, resulting from the existence of increasing returns to scale (reflected in the number of varieties of goods produced) in the sector of manufactured products and transport costs, which creates and supports the circular and cumulative processes. Because, there are transportation costs and the immobility of resources from agriculture, there are centrifugal forces that oppose the centripetal forces of agglomeration. These authors attempt to explain, too, how circular and cumulative process (which leads to agglomeration) begins, in other words, considering an initial situation of equilibrium, the process begins with shocks disturbing the stable situation that come mainly from the difference in wages real, between the two locations.

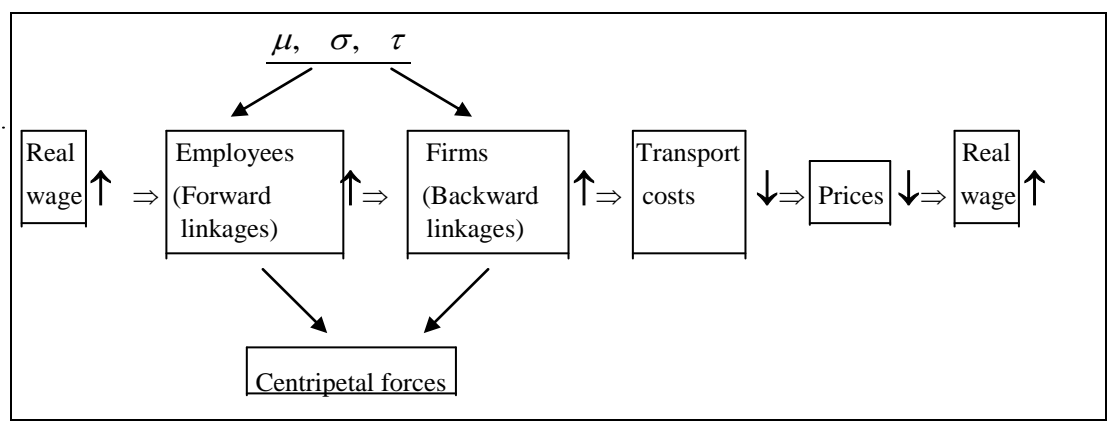

***Figure 2:*** *A mechanism that describes the process of agglomeration*

In this context, it appears that between the two approaches there are some similarities, such as those mentioned earlier, but there are many differences, such as the fact the authors associated with the agglomeration process ignore, at least explicitly, technical progress, productivity, human capital, the accumulation of capital, efficiency wages and infrastructure, which is understandable given the concerns and objectives and limits of this approach to build tractable models. Also, ignore the institutions and policies, a fact which is admitted by these authors, because there are few empirical studies that support their definition safely. Ignore, too, the forces of demand, given the reference to real wages and "backward and forward" linkages. Finally, do not consider the industry as the only sector with increasing returns, but all sectors that produce differentiated products. Besides the above, for some authors, the work of the agglomeration has some less good aspects, such as the fact that they are works based on economic principles few rigorous ((15)Nijkamp, 2001); make a distinction not good between the economic theory (without define its precise meaning) and the economy of the location (Nijkamp 2001); exaggerate the lack of references to work in these areas, which is not true, especially in the last three decades, especially in the regional economy (Nijkamp 2001); exaggerate in the criticism of the "Central Place Theory" of Christaller, when referring that is a classification scheme, this taking into account the works that have been developed in this field (Nijkamp, 2001); lack of conceptual precision when is considered the modern agricultural sector without increasing returns; the center-periphery concept is not well defined, etc. (Nijkamp, 2001), and the mathematical rigor and theoretical integration are primary ((16)Pavlik, 2000). In summary, these publications on the one hand, seek to establish a new parameterization and a new strategy for modeling spatial models of the economy and promote it as the New Economic Geography, but may not be, says Pavlik (2000). This is because the theoretical orthodoxy of the approach and the use of strategies to consider simplifications to analyze the real world, ignores many of the issues of central interest to heterodox economists, including evolutionary and institutional economists and economic geographers. So, while represents a contribution for some disciplines and bring the economic models of spatial economics theory to the heart of neoclassical economics, the approach outlined can hardly be argued to serve as a substitute for the issues, interests and approaches of contemporary economic geographers (Pavlik, 2000) . On the other hand, it may be noted that these works are an effort to approach the general economic theory and the spatial economy with some interesting ideas and input. However, seeking to build a bridge based on concepts few simples (and certainly not universally accepted), in particular the model of imperfect competition, marked by increasing returns to scale (from Dixit and Stiglitz). They admit, too, the emergence of various types of spatial agglomeration models, in open system (multi-region and multi-country), taking into account transportation costs, "forward and backward" linkages and immobility of resources. However, any exposure about the advantages of agglomeration has to start in the basic economic principles, developed by several investigators in polar growth theories, growth center or pole of attraction. It should be added that the two pillars of the regional economy are certainly agglomeration economies and transportation costs. But the emphasis that is placed in publications, the analysis of urbanization and economic advantages of scale, compared with what is given to transport costs, is certainly disproportionate. In addition, the reference to costs, "iceberg" of Samuelson, may be insufficient to explain the emergence of the global urban scheme. The authors of these publications refer to nothing, too, the theory of equilibrium and the price space, rigorously developed by several regional economists. Nothing concern also of the recent endogenous growth theory (Nijkamp, 2001). Capital mobility, foreign ownership and foreign control of production are barely mentioned. Only the industrial life cycle and demand of resources for reasons of foreign investment are mentioned, while others such as the power market and the internationalization of business are left out ((17)Jovanovic, 2000). As a final note, be noted that, indeed, these works of Fujita, Krugman and Venables is a good contribution to the systematization of matters related to economic geography, especially with attempts to model these issues and to bring the Space Economy Economy to the economic theory. Surely, has many fail and the attempt to call these contributions of New Economic Geography can at this stage be overstated, not least because the empirical work in these areas are scarce, however, open a new field of research to be highly relevant.

The authors of polarization always try, where is possible, avoiding artificial assumptions, ignoring transport costs (although Kaldor refers to them in a subtle way). Ignore, too, questions of location, although there are references to these aspects from Young (1928), for example, reinforced by (18)Thirlwall (1999), perhaps because these issues are present in all current economic, such as, apart that we have been referring, the endogenous growth theory, largely associated with Romer and Lucas, among others.

(19)Romer et al. (1999), for example, says that lately has been looking at part of international trade, as determined by geographical factors. That is, some countries trade more because they are near well-populated countries. Changes in trade that are due to geographical factors, may serve as a natural test to identify the effects of trade. The test results are consistent and show that in this way, trade increases income. The income increases because trade increases the accumulation of physical and human capital and because it increases the output for a given level of capital. The same is true in domestic and international trade.

The authors of polarization ignore, yet, the immobility of certain resources (although admit it, despise it, because they consider the industry as the engine of economic development). Besides the above, in addition, Myrdal (1957), for example, in the exposition about the causes of the circular and cumulative processes seems to try include a key role for the economies of scale. But (20)Krugman (1995) states that it was unable to find a single reference to their role, even indirect, in this work. On the other hand, when Myrdal (1957) offers an example of the causes of circular process, the external economies occur via tax rates, rather than by the market of some "spillover". Myrdal's central thesis was the idea of circular and cumulative causes. However, it is worth noting that these ideas of circular and cumulative causes were already referred in Alyn Young (1928), for not mention (21)Rosenstein-Rodan (1943) and Nurkse. Myrdal, in fact, organized an extensive and familiar set of ideas. (22)Hirschman (1958) also used the idea of linkages, although it was more distinct in terms of efficacy and the policies that derive from there, was little in the new-intellectual, because, in reality, Rosenstein-Rodan (1943) have had spoken of linkages and (23)Fleming (1955) had already been explicitly, when mention to the "backward and forward" linkage, among others. Krugman (1995) refers also lacking microeconomic foundation for the work of polarization.

## 3. CONCLUSIONS

From the above analysis we see that both approaches seek to explain the concentration of population and economic activity in certain places, but with different perspectives, ie the polarization optical in economic development and agglomeration in a spatial perspective.

The explanations of the two processes are based on the theoretical work of the "High Development Theory", especially in the publications of Myrdal (1957) and Hirschman (1958), although these have been influenced by others before them who stress aspects of increasing returns to scale, the cumulative processes and endogenous factors of production, in contrast to neoclassical theorists who admit the existence of constant returns (or decreasing), stable equilibria and exogenous resources (because it admitted the development guided by the forces of supply of inputs).

However, despite these few similarities, the two processes have a number of differences, even as they followed different ways, resulting from having different goals and concerns.